\documentclass{article}

\usepackage{PRIMEarxiv}
\usepackage{tabularx}
\usepackage[utf8]{inputenc} 
\usepackage[T1]{fontenc}    
\usepackage[hidelinks,urlcolor=blue]{hyperref}
\usepackage{url}            
\usepackage{booktabs}       
\usepackage{amsfonts}       
\usepackage{nicefrac}       
\usepackage{microtype}      
\usepackage{lipsum}
\usepackage{fancyhdr}       
\usepackage{graphicx}       
\graphicspath{{}}     

\title{The AI Community Building the Future? A Quantitative Analysis of Development Activity on Hugging Face Hub}

\author{
  Cailean Osborne\thanks{Corresponding author: Cailean Osborne, cailean.osborne@oii.ox.ac.uk} \\
  Oxford Internet Institute \\
  University of Oxford \\
  Oxford, UK\\
  \texttt{cailean.osborne@oii.ox.ac.uk} \\
   \And
  Jennifer Ding \\
  The Alan Turing Institute \\
  London, UK\\
  \texttt{jding@turing.ac.uk} \\
 \AND
  Hannah Rose Kirk \\
  Oxford Internet Institute \\
  University of Oxford \\
  Oxford, UK\\
  \texttt{hannah.kirk@oii.ox.ac.uk}
}

\begin{document}
\maketitle
\vspace{-1em}

\begin{abstract}
Open model developers have emerged as key actors in the political economy of artificial intelligence (AI), but we still have a limited understanding of collaborative practices in the open AI ecosystem. This paper responds to this gap with a three-part quantitative analysis of development activity on the Hugging Face (HF) Hub, a popular platform for building, sharing, and demonstrating models. First, various types of activity across 348,181 model, 65,761 dataset, and 156,642 space repositories exhibit right-skewed distributions. Activity is extremely imbalanced between repositories; for example, over 70\% of models have 0 downloads, while 1\% account for 99\% of downloads. Furthermore, licenses matter: there are statistically significant differences in collaboration patterns in model repositories with permissive, restrictive, and no licenses. Second, we analyse a snapshot of the social network structure of collaboration in model repositories, finding that the community has a core-periphery structure, with a core of prolific developers and a majority of isolate developers (89\%). Upon removing the isolates from the network, collaboration is characterised by high reciprocity regardless of developers' network positions. Third, we examine model adoption through the lens of model usage in spaces, finding that a minority of models, developed by a handful of companies, are widely used on the HF Hub. Overall, activity on the HF Hub is characterised by Pareto distributions, congruent with OSS development patterns on platforms like GitHub. We conclude with recommendations for researchers, companies, and policymakers to advance our understanding of open AI development.
\end{abstract}


\section{Introduction}
Open source developers have become central actors in the political economy of artificial intelligence (AI). The rise of open source AI, specifically the practice of releasing and fine-tuning pre-trained models that are freely available\footnote{N.B. We use ``open models'' because they do not meet the requirements to be considered ``open source'' \cite{osi_open_2024,osi_open_2007}.}, has extended open science practices crucial to AI advances \cite{langenkamp_how_2022,white_model_2024}, including open source software (OSS)\footnote{As defined by the OSI \cite{osi_open_2007}, OSS is software source code that anyone can inspect, use, modify, or redistribute.} and open access to research papers \cite{arxiv_arxivorg_2024} and datasets \cite{kaggle_find_2024, commoncrawl_common_2024,imagenet_imagenet_2024}. Open source AI has attracted attention as a potential challenger to the dominance of startups and Big Tech in AI research and development (R\&D) \cite{ahmed_growing_2023, tarkowski_mirage_2023}. Grassroots initiatives like EleutherAI \cite{eleutherai_eleutherai_2021}, BigScience \cite{akiki_bigscience_2022}, and BigCode \cite{huggingface_bigcode_2024} have shown the feasibility of open model development \cite{ding_towards_2023}, while the Hugging Face (HF) Hub has emerged as a popular platform used by millions to host, download, and collaborate on a growing number of models, datasets, and spaces (i.e., web applications to demonstrate and try out models) \cite{huggingface_hugging_2024}.

While the benefits and risks of open source AI have been widely debated \cite{law_open-source_2023,  solaiman_gradient_2023,kapoor_societal_2024, seger_open-sourcing_2023,eiras_risks_2024}, the practices and processes involved in open model development have received relatively little attention. To date, only a handful of scholars have explored various aspects of open model development, including user contributions to grassroots initiatives \cite{akiki_bigscience_2022,ding_towards_2023}, commercial participation in model development \cite{ding_towards_2023,widder_open_2023}, model maintenance \cite{castano_analyzing_2024}, and the processes and tools used by adjacent open data engineering communities \cite{heltweg_systematic_2023}. 

We contribute to this nascent research agenda with a three-part quantitative analysis of development activity on the HF Hub. First, we investigate typical patterns of various types of activity on the HF Hub in 348,181 model, 65,761 dataset, and 156,642 space repositories (\textbf{RQ1}). 
Subsequently, we apply social network analysis (SNA) of code contributions to model repositories to investigate the social network structure of the developer community as well as collaboration practices amongst developers (\textbf{RQ2}). We replicate this analysis for models in the sub-fields of natural language processing (NLP), computer vision (CV), and multimodal (MM) for comparative analysis. 
Finally, we quantify model adoption through the lens of model usage in spaces on the HF Hub (\textbf{RQ3}), providing insights into the widespread use of a minority of models in the HF Hub developer community and the key actors driving their development. 

Overall, our analysis reveals that various aspects of development activity on the HF Hub---e.g., interactions in model, dataset, and space repositories; collaboration in model repositories; and model adoption in spaces---exhibit right-skewed, Pareto distributions, which is a well-documented pattern in OSS development \cite{goeminne_evidence_2011,mockus_two_2002,szymanski_applicability_2023,xu_12_2006,zhang_companies_2021}. While the open model development life-cycle involves unique practices which differ from OSS development \cite{castano_analyzing_2024}, such as model training and fine-tuning, the observed similarities in the overall patterns of activity suggests that future research on open source AI can benefit from drawing on the extensive, multidisciplinary literature on the social dynamics of OSS development. Based on our findings, we propose a number of recommendations for researchers, policymakers, and platform providers to facilitate research and evidence-based discussions on open source AI.

The paper has the following structure. First, the literature review provides an overview of prior work on open source AI, as well as prior work on OSS development in order to draw comparisons between OSS and open model development practices. Second, we presents the RQs and research design. Third, we introduce the main findings from the three-part analysis. Fourth, we discuss the contributions of the findings to research and practice, and make recommendations for research and practice. We conclude with a discussion of what further clarification of the practices in open model development can offer for (open source) AI researchers, developers, policymakers, and platform providers.

\section{Related Work}

\subsection{``We Have No Moat'': The Emergence of Open Models}

Open science practices, from the development of open source software (OSS) to the provision of open access to research papers (e.g., via arXiv \cite{arxiv_arxivorg_2024}) and datasets (e.g., via Kaggle \cite{kaggle_find_2024}, ImageNet \cite{imagenet_imagenet_2024}, or Common Crawl \cite{commoncrawl_common_2024}), have been integral to advances in AI R\&D and adoption \cite{white_model_2024,paperswithcode_papers_2023}. The culture and norms of openness in AI has evolved significantly in the last 15 years \cite{gururaja_build_2023}. For example, in 2007, a coalition of 16 researchers lamented the lack of OSS that standardised the implementation of ML algorithms, highlighting this as a major obstacle to advances and reproducibility in AI research \cite{sonnenburg_need_2007}. Yet today AI R\&D is simply unimaginable without OSS \cite{langenkamp_how_2022,osborne_public-private_2024}, drawing on a growing commons of over 300 OSS libraries \cite{haddad_artificial_2022}, hundreds of thousands of open models \cite{huggingface_transformers_2023}, and over a million OSS repositories \cite{github_machine_2023}.  

Following years of debate about the safety of openly releasing AI models \cite{solaiman_release_2019,solaiman_gradient_2023,kapoor_societal_2024,bender_dangers_2021}, recent years have seen the emergence and proliferation of ``open'' models, which individuals and organisations have shared on an open access basis on platforms such as the HF Hub \cite{white_model_2024}. Prior to this, AI models, in particular large language models (LLMs), were principally developed and maintained behind closed doors, albeit with open science practices, such as the sharing of publications on arXiv and code on platforms like GitHub. The start of this trend is attributed to EleutherAI, a grassroots research group, which formed on a Discord server with the intention to develop and release an open source variant to OpenAI’s GPT, resulting in \texttt{The Pile} in December 2020 \cite{gao_pile_2020}, a library of datasets for training LLMs, and \texttt{GPT-Neo} in March 2021 \cite{black_gpt-neox-20b_2022}. Subsequently, open models gained more visibility with the release of other state-of-the-art AI models \cite{tarkowski_mirage_2023}, including \texttt{BLOOM} by the BigScience workshop in July 2022 \cite{bigscience_workshop_bloom_2023}, \texttt{Stable Diffusion} by Stability AI in August 2022 \cite{stability_ai_stable_2022}, and \texttt{LLaMA 2} by Meta in July 2023 \cite{meta_meta_2023}, amongst others. 

The proliferation of open models, especially foundation models, has ignited heated debate about their potential benefits and risks \cite{kapoor_societal_2024,law_open-source_2023,solaiman_gradient_2023,bdeir_introducing_2024, seger_open-sourcing_2023,eiras_risks_2024}. On the one hand, open models are said to promise benefits for research, innovation, and competition by lowering entry barriers and widening access to state-of-the-art AI \cite{cihon_helping_2024}. Drawing on Linus' Law from OSS development that ``given enough eyeballs, all bugs are shallow'' \cite{raymond_cathedral_2001}
, proponents argue that open model development and auditing offers safety advantages \cite{wladawsky-berger_are_2023}. 
In addition, open access to models lowers the barriers for adaptability and customisation for diverse language contexts \cite{pipatanakul_typhoon_2023,kapoor_societal_2024}. On the other hand, open models can pose risks of harm by both well-intended and malicious actors, including the creation of deepfakes \cite{nguyen_deep_2022,lakatos_revealing_2023,thiel_generative_2023}, disinformation \cite{goldstein_generative_2023,musser_cost_2023}, and malware \cite{tsamados_cybersecurity_2023, david_chatgpt_2023}. A study by 25 experts concluded that open models have five distinctive properties that present \textit{both} benefits and risks: broader access, greater customisability, local adaptation and inference ability, the inability to rescind model access, and the inability to monitor or moderate model usage \cite{kapoor_societal_2024}.

The development of open models has been described as a potential challenge to the dominance of Big Tech companies in AI R\&D \cite{ahmed_growing_2023, gulson_steering_2021}. This was underlined by a leaked Google memo that claimed there is ``no moat around closed-source AI development'' and ``open source solutions will out-compete companies like Google or OpenAI'' \cite{patel_google_2023}. Venture capitalists have bullishly invested in open source AI startups \cite{wiggers_5_2023, abboud_mistral_2024}, and world leaders like President Macron of France have pledged public funds to support open source AI \cite{chatterjee_france_2023}. In addition, the Mozilla Foundation has launched mozilla.ai with \$30 million in investment to build a trustworthy, independent, and open source AI ecosystem ``outside of Big Tech and academia'' \cite{mozilla_foundation_introducing_2023}. While proponents champion open models as good news for innovation and competition, others temper this optimism by pointing to market concentrations at several layers of the AI stack, from chips to cloud compute infrastructure, which remain unchallenged by innovations stemming from open source AI communities \cite{lehdonvirta_cloud_2023,srnicek_data_2022,widder_open_2023}.

A myriad of meanings are attached to ``open models'' and ``open source AI''. Oftentimes these terms are understood as making pre-trained models, parameters (or ``weights''), and documentation available on platforms like the HF Hub. In some cases, they refer to open collaboration on the development of models \cite{ding_towards_2023}. The description of open models as ``open source'' has been fiercely contested for failing to meet OSS standards as defined by the OSI \cite{osi_open_2007,white_model_2024,maffulli_metas_2023,nolan_llama_2023}. For example, when Meta imposed limits on use of LLaMA 2, Stefano Maffulli from the OSI commented, ``Unfortunately, the tech giant has created the misunderstanding that LLaMA 2 is `open source' – it is not. Meta is confusing 'open source' with `resources available to some users under some conditions,' [which are] two very different things'' \cite{maffulli_metas_2023}. 

Companies have been criticised for ``open-washing'' by promoting their models as ``open source'' models, when they are typically ``open weight'' models at most, as a commercial strategy to present themselves as patrons of the digital commons, whilst disguising their intent to set open standards and benefit from crowdsourced innovation \cite{widder_open_2023,srnicek_data_2022, liesenfeld_rethinking_2024, liesenfeld_opening_2023}. A review of the openness of LLMs found that, ``[W]hile there is a fast-growing list of projects billing themselves as 'open source', many inherit undocumented data of dubious legality, few share the all-important instruction-tuning (a key site where human annotation labour is involved), and careful scientific documentation is exceedingly rare'' \cite{liesenfeld_opening_2023}. 

It remains an open question whether one can or should classify AI models as either open or closed-source. Through a global, multi-stakeholder approach, the OSI is currently developing a definition of open source AI as AI systems that are made available under terms that grant the freedoms to use, study, modify, and share the system \cite{osi_deep_2023,osi_open_2024}. Countering binary approaches, Irene Solaiman \cite{solaiman_gradient_2023} makes the case that AI systems are not either fully open or fully closed; rather, the openness of AI systems can be plotted along a gradient with six degrees of openness. Each grade of openness involves trade-offs between concentrating power and mitigating risks \cite{solaiman_gradient_2023}. As the field rapidly evolves, developing responsible practices, norms, and regulation around open source AI remains a critical challenge \cite{bdeir_introducing_2024,cihon_helping_2024}. 

\subsection{A Nascent Research Agenda on Open Source AI}

While the benefits and risks of open models have been widely discussed, we still have a limited understanding of the collaborative practices involved. In this section, we review prior work on open model development and motivate our empirical analysis of development activity on the HF Hub to address this research gap.

The HF Hub has emerged as a popular platform used by individuals and organisations to share, download, and collaborate on models, datasets, and spaces \cite{huggingface_hugging_2024-1,ait_suitability_2023}. The HF Hub is a ``model marketplace,'' which is ``a new form of user-generated content platform, where users can upload AI systems and AI-related datasets, which in turn can be downloaded, and depending on the business model, queried, tweaked, or built upon by other users'' \cite{gorwa_moderating_2024}. Much of the activity amongst the emerging developer community on this platform concerns individuals fine-tuning pre-trained models that were released by industry leaders for downstream use in research and applications \cite{widder_open_2023}.  In addition to the hosting and fine-tuning of open models, a few grassroots initiatives have embraced open collaboration methods to develop open models. For example, the development of BLOOM, a 176B parameter multilingual LLM, and its training dataset, ROOTS, was the largest ``open source'' AI collaboration to date, involving over 1,000 volunteers from over 70 countries and over 250 institutions \cite{akiki_bigscience_2022}. Such initiatives have demonstrated alternative pathways for AI model development beyond the handful of companies that dominate the AI R\&D \cite{ahmed_growing_2023,ding_towards_2023}. Prior work has also highlighted the leadership role of companies, such as Hugging Face, in organising ``values-driven initiative[s]'', such as the BigScience workshop, and attracting contributors who have diverse motivations, from developing new skills and working on new problems to publishing research giving back to the ecosystem \cite{ding_towards_2023,akiki_bigscience_2022}. 

Due to the growing popularity of the HF Hub, scholars have examined the suitability of the HF Hub for empirical research on open model development \cite{ait_suitability_2023,ait_hfcommunity_2023}.\footnote{The authors define ``suitability for empirical research'' as ``the amount and adequacy of the features to enable software development practices and the sufficient quantity of data to enable the conduction of empirical studies about such practices'' \cite{ait_suitability_2023}.} Castaño et al. \cite{castano_analyzing_2024} provide most comprehensive empirical insights into maintenance practices in model repositories on the HF Hub.\footnote{Model maintenance is defined as ``a higher number of commits, regular commit frequency, shorter intervals between commits, fewer days without commits, and a slightly higher number of authors'' \cite{castano_analyzing_2024}.} They find that commit activity follows a right-skewed distribution, with a few models receiving extensive activity while the majority of repositories receive limited activity \cite{castano_analyzing_2024}. While the majority of models are developed by singular developers (1.18 mean, 1.0 median), some model repositories, such as $bigscience/bloom$ or $bigcode/santacoder$, are co-developed and co-maintained by up to 20 developers \cite{castano_analyzing_2024}. They also find that developers tend to prioritise ``perfective tasks'' to enhance model performance and align with technological advances, unlike OSS maintenance that focuses on bug fixes and feature additions \cite{castano_analyzing_2024}. The authors contend this ``reveals the need for methods and tools specifically designed for the unique demands of ML model maintenance. Such tools may include advanced version control systems optimized for data and model tracking, as well as automated monitoring tools capable of detecting model drift or degradation'' \cite{castano_analyzing_2024}. Prior work has also examine carbon emission reporting in model repositories, finding stagnation in emissions reporting by developers and highlight and the need for improved reporting practices and carbon-efficient model development on the HF Hub \cite{castano_exploring_2023}.

Our research builds on this prior work. As one of the first studies to investigate open model development practices, in the next section we draw on prior work on OSS development in order to be able to compare our findings to prior research and to lay the groundwork for a more comprehensive understanding of open model development in the future.

\subsection{Learning from Prior Work on OSS development}\label{HF-litreview-OSS}

Prior work on OSS development provides a valuable framework for understanding the social dynamics of open model development. In the early 2000s, a number of metaphors were used to describe the social structure of ``the OSS community''. For example, the Linux developer community was described as a ``bazaar'' that vibrated with the activity of geeks, hackers, and hobbyists, who performed various tasks, from bug-spotting to writing code to ``serving the hacker culture itself'' \cite{raymond_cathedral_2001}. However, prior work illustrates that OSS communities have diverse social structures \cite{eghbal_working_2020,zhou_inflow_2016}, from ``caves'' with singular developers \cite{krishnamurthy_cave_2005} to ``core-periphery'' networks, akin to ``layered onions'' \cite{crowston_effective_2005}, with uneven activity distributions ranging from core contributors (e.g., project initiators) to users (e.g., bug-spotters) \cite{bird_mining_2006,crowston_hierarchy_2006,long_social_2007,orucevic-alagic_network_2014}.

Numerous studies highlight that various types of activity in OSS development, such as discussions in mailing lists, bug-spotting in issue trackers, and commit activity, exhibit right-skewed, Pareto distributions \cite{goeminne_evidence_2011,szymanski_applicability_2023,zhang_companies_2021}. Indeed, it is well-documented observation that OSS development is typically characterised by the Pareto principle, commonly known as the 80/20 rule or the law of the vital few, which states that approximately 80\% of effects come from 20\% of causes \cite{wood_joseph_2005}. These findings are congruent with a wide range of Internet phenomena, which similarly exhibit right-skewed distributions, which follow power laws \cite{faloutsos_power-law_1999,mahanti_tale_2013}. However, there are exceptions to the rule; for example, a study of 2,496 projects on GitHub found that the Pareto principle does not always characterise development activity in OSS repositories, thus highlighting the need to be cautious about generalising the Pareto principle as an incontestable law of OSS development \cite{yamashita_revisiting_2015}. Furthermore, many activities, such as mentorship and hackathons, take place outside of the repository \cite{geiger_labor_2021,hossain_regional_2021,osborne_public-private_2023,takhteyev_coding_2012} and are therefore invisible to quantitative scholars of OSS development practices.

The various social structures of OSS communities are shaped, amongst others, by the diverse incentives of individuals and companies that participate in OSS development \cite{feller_understanding_2002,bonaccorsi_comparing_2006,li_systematic_2024}. Individual developers are typically motivated by factors such as personal values, altruism, enjoyment, reputation-building, and career benefits \cite[]{von_krogh_carrots_2012,shah_motivation_2006,lakhani_why_2003,ghosh_freelibre_2002}. However, there are also major barriers to participation, including gender disparities \cite{brooke_trouble_2021,vasilescu_gender_2014} and geographic inequalities \cite{hossain_regional_2021,takhteyev_coding_2012}. Activity tends to be concentrated in the Global North \cite{braesemann_global_2019} and the English \textit{lingua franca} is a barrier for many developers \cite{takhteyev_coding_2012,williams_enabling_2023}. Furthermore, the incentives of OSS developers vary by geography: while developers in the USA show a relatively strong interest in ``geek culture'', developers in India and China tend to be motivated primarily by career benefits \cite{subramanyam_freelibre_2008}. Thus, ``researchers studying open source should be mindful of geographic variation in what motivates participation and what forms participation may take, particularly outside of the code repository'' \cite{hossain_regional_2021}.

Meanwhile companies primarily participate in OSS development for strategic reasons, such as recruiting developers \cite{agerfalk_outsourcing_2008,birkinbine_incorporating_2020, west_challenges_2006}, reducing costs \cite{chesbrough_measuring_2023,lindman_beyond_2009,birkinbine_incorporating_2020}, influencing OSS projects \cite{dahlander_man_2006,lindman_beyond_2009}, promoting open standards \cite{chesbrough_measuring_2023,fink_business_2003,lerner_simple_2002}, and building a reputation as an OSS patron \cite{bonaccorsi_comparing_2006,pitt_penguins_2006, osborne_public-private_2024}. Commercial participation has mixed effects on the social structure of OSS communities. Typically, one company or a few companies emerge as dominant contributors in projects \cite{zhang_companies_2021,nguyen-duc_software_2019}. The dominance of a company is negatively associated with the participation of volunteers, while it is positively associated with the productivity of contributors and the quality of issue reports \cite{zhang_corporate_2022,zhou_inflow_2016}. It is also common for companies, which may be market rivals, to collaborate in OSS ecosystems \cite{germonprez_open_2013,linaker_how_2016,nguyen-duc_software_2019, teixeira_collaboration_2014,zhang_how_2020}, which has turned many OSS communities ``from networks of individuals into networks of companies'' \cite{agerfalk_outsourcing_2008}. 

Building on this prior work, this study aims to provide novel insights into the collaborative dynamics of the HF Hub. Specifically, we investigate typical patterns of development activity across model, dataset, and space repositories the HF Hub (\textbf{RQ1}), the social network structure of its developer community (\textbf{RQ2}), as well as adoption and key actors driving the development of the most widely-adopted models (\textbf{RQ3}). The research extends the literature by shedding light on the practices involved in open model development on this increasingly important platform. The findings contribute to a more comprehensive understanding of and lay the groundwork for future research on open model development.

\section{Study Design}

\subsection{Research Aims \& Research Questions}
This study extends the nascent research agenda on open model development with a quantitative analysis of development activity on the HF Hub. We adopted a quantitative approach to explore large-scale patterns and trends in development activity on the HF Hub, which is a suitable approach when one seeks to generate baseline insights on a new phenomenon \cite{easterbrook_selecting_2008}. In particular, we examine different aspects of development activity on the HF Hub via the following RQs:

\begin{itemize}
    \item \textbf{RQ1}: What are typical patterns of development activity across the HF Hub?
    \item \textbf{RQ2}: What is the social network structure of the HF Hub developer community?
    \item \textbf{RQ3}: What is the distribution of model adoption on the HF Hub, and who are the key actors driving the development of the most widely-adopted models?
\end{itemize}

These RQs examine different aspects of development activity on the HF Hub. \textbf{RQ1} focuses on identifying common patterns across various types of activity, such as likes, discussions, commits, and downloads, in the repositories of models, datasets, and spaces. Concretely, this analysis expands prior work that focuses on commit activity in model repositories \cite{castano_analyzing_2024}. \textbf{RQ2} concerns the social network structure of the developer community on the HF Hub. In particular, we analyse a snapshot of collaboration interactions in model repositories amongst around 100,000 developers, building on prior descriptions of collaboration on open models \cite{ding_towards_2023, akiki_bigscience_2022} and maintenance practices \cite{castano_analyzing_2024}. Lastly, \textbf{RQ3} empirically tests a prior observation of uneven model adoption and the influence of Big Tech companies \cite{widder_open_2023} by examining the distribution of model use in spaces and identifying the developers of the most used models. In addition, we examine model co-usage patterns to provide insights into the interdependencies and ecosystems surrounding popular models.

\subsection{The HF Hub: A New Platform \& Source of Research Data}
The HF Hub was launched in 2021 by Hugging Face, a startup whose mission is to ``democratize AI'' \cite{huggingface_hugging_2024-1}. The HF Hub is a Git-based social coding platform, widely used by researchers, developers, and hobbyists to share, discover, discuss, and collaborate on open models \cite{huggingface_models_2024}, datasets \cite{huggingface_datasets_2024}, and spaces \cite{huggingface_spaces_2024}. Spaces are interactive web applications that facilitate the creation of demonstrations and make models hosted on the platform more accessible to end-users. The platform provides a number of tools for open model development, such as version control for collaboration and tracking \cite{huggingface_models_2024}, and evaluation and benchmarking of model performance \cite{huggingface_evaluate_2024}. The HF Hub API allows programmatic access to platform resources as well as metadata from repositories hosted on the platform \cite{huggingface_hugging_2024}. In light of its features and data availability, prior work underlines the platform's suitability for empirical studies on open model development \cite{ait_suitability_2023,castano_analyzing_2024}. Building on this prior work, this paper aims to advance the research community's understanding of the development practices in open model development as well as methodological considerations regarding the HF Hub. 

When using data from the HF Hub, it is important to consider the ethical implications and adhere to the platform's terms of service. In the study, we only collected publicly available data through the official HF Hub API, respecting the privacy settings of users and repositories. For example, we did not attempt to access or include data from private repositories in the analysis. Additionally, we anonymised the collected data by focusing on aggregate measures and avoiding the disclosure of personally identifiable information in the findings. Ethical clearance for this study was obtained from the CUREC institutional review board at the University of Oxford.

\subsection{Data Collection}\label{HF-datacollection}
We collected data via the HF Hub's API in October 2023 \cite{huggingface_hugging_2024}, using Python scripts that are available on GitHub \cite{osborne_python_2024}. For \textbf{RQ1}, we collected and processed metadata for a number of activities from the public repositories of 348,181 models, 65,761 datasets, and 156,642 spaces, using the \texttt{list\_models()}, \texttt{list\_datasets()}, and \texttt{list\_spaces()} API endpoints. These included: likes (\texttt{n\_likes}), downloads (\texttt{n\_downloads})\footnote{N.B. We do not report data for downloads of spaces because spaces cannot be downloaded.}, discussions (\texttt{n\_discussions}), commits (\texttt{n\_commits}), unique developers who have contributed commits (\texttt{n\_commiters}), unique developers who started discussions (\texttt{n\_disc\_starters}),\footnote{N.B. As per the API, data collection for participation in discussions was limited to users that had started discussions. It was not possible to collect data about users that had made comments in discussion threads.} and the repository's community size (\texttt{n\_community}), calculated as the cardinality of the set union of \texttt{n\_disc\_starters} and \texttt{n\_commiters}. As per prior work \cite{lin_developer_2017,zhang_how_2020,robles_developer_2005}, we removed bots and merged multiple developer identities before enumerating \texttt{n\_disc\_starters}, \texttt{n\_commiters}, and \texttt{n\_community}. As a result, \texttt{n\_community} is recorded as 0 if no user has made a commit or started a discussion in the repository, which ignores the creator of the repository. We acknowledge that alternatively such repositories could have the value 1.

For \textbf{RQ2}, we operationalised collaboration on models as instances where a pair of developers contributed commits to the same model repository, with  direct edges recorded between developers that were weighted by the number of times a developer contributed a commit to the same repository as the other developer \cite{lopez-fernandez_applying_2004}. We operationalised commit activity as acts of collaboration because commits are easily measurable, represent ``validated'' contributions, and represent an accurate audit trail of collaboration \cite{orucevic-alagic_network_2014, zhang_how_2020}. However, we acknowledge that the fact that two developers commit to the same repository does not necessarily imply direct interaction; for example, it would have been more accurate to focus on developers' contributions to the same file in a repository, as we discuss in Section \ref{HF-threats-to-valdity}. Formally, we modelled collaboration as a network $N = (D,E,W)$, where $D$ is the set of developers, $E = \{(i,j,w_{ij}) \mid i,j \in D, w_{ij} \in \mathbb{N}\}$ is the set of directed edges denoting the relationships between developers, and $W = \{w_{ij} \mid (i,j,w_{ij}) \in E\}$ represents the weights associated with each directed edge. For a developer pair $i$ and $j$, we denote the directed relationship as $(i,j,w_{ij})$, where $w_{ij}$ signifies the number of times developer $i$ has committed to the same repository as developer $j$.

To collect data for the analysis of \textbf{RQ2}, we collected commit data from public model repositories via the HF Hub API. We started by retrieving a list of all available model IDs using the \texttt{list\_models()} endpoint. Then, for each model repository, we used the \texttt{list\_repo\_commits()} endpoint to retrieve the commit data, including the authors associated with each commit. For each commit, we recorded an edge between the the developer who made the commit (\texttt{source\_node}) and all other developers who had contributed to the repository (\texttt{target\_node}). In cases where a repository had only one contributor, we created self-loop edges to capture the isolate contributor's activity. We did not take temporal dynamics of commit activity into account, which we discuss as a threat to construct validity under Section \ref{HF-threats-to-valdity}. We collected data for collaboration in NLP, CV, and MM model repositories by filtering repositories based on the tags, which developers add to their repositories to aid discoverability on the HF Hub. We used the list of tags per sub-field provided by the HF Hub, including  \texttt{computer-vision} and \texttt{image-classification} for CV models; \texttt{translation} and \texttt{summarization} for NLP models; and \texttt{image-to-text} and \texttt{image-to-video} for MM models.

For \textbf{RQ3}, we collected data on model usage in spaces using the \texttt{list\_models()} and \texttt{model\_info()} API endpoints. We modelled model usage in spaces as a bipartite network, akin to the representation of software dependency networks \cite{savic_complex_2019}. The bipartite model usage network is denoted as $D = (M, S, E)$, where $M$ is the set of models, $S$ is the set of spaces, and $E = \{(m, s) \mid m \in M, s \in S\}$ is the set of undirected edges signifying that ``space'' $s$ uses model $m$. The edges are unweighted, representing the model usage relationship between a ``space'' and a model. From the bipartite network $D$, we derived an undirected model co-usage network $C = (M, E, W)$. In this network, $M$ is the set of models, $E = \{(m_i, m_j) \mid m_i, m_j \in M\}$ is the set of undirected edges connecting models based on their co-usage in a ``space'', and $W = \{w_{ij} \mid (m_i, m_j) \in E\}$ is the set of weights assigned to the edges, reflecting the frequency of co-usage of models $m_i$ and $m_j$ across spaces. This analysis complements the former analysis of model usage with insights into the interdependencies and ecosystems surrounding widely used models on the HF Hub.

\subsection{Username Merging} \label{HF-dataprocessing}
Following prior work, before the analysis, we undertook data preprocessing to merge multiple developer identities per unique developer, which can be caused by how Git records usernames based on users' local repository credentials \cite{bird_mining_2006,goeminne_comparison_2013,kouters_whos_2012,robles_developer_2005,zhang_companies_2021}. We assumed this might be an issue on the HF Hub, too. To ensure the accuracy of the dataset of 101,144 developers, we applied a three-pronged approach. First, we classified username string similarity (threshold=90\%) between pairs of developers who contributed to the same repository, accepting 126 out of 180 (70.00\%) pairs based on manual username searches on the HF Hub. Second, in light of the presence of potential real names (i.e. usernames with spaces like ``Jessica Smith''), we examined string similarity (threshold=90\%) between 1,979 potential real names and the remaining 99,041 usernames, accepting 358 out of 403 (87.75\%) username pairs after manual searches on the HF Hub. Finally, we inspected the usernames of 700 developers with a network degree of 10 or higher, who represented 0.7\% of developers but accounted for 44.78\% of edges, via manual searches on the HF Hub. This resulted in the identification of 212 username pairs. In total, we merged 546 usernames after removing duplicates.

\subsection{Data Analysis}
To investigate development activity on the HF Hub (\textbf{RQ1}), we conducted a descriptive analysis of various types of activity in 348,181 model repositories, 65,761 dataset repositories, and 156,642 space repositories. Pearson correlation coefficients were calculated to assess the pairwise relationships between the activity variables. In addition, we employed the Mann-Whitney \textit{U} test to compare the levels of activity across repositories with different licenses (Permissive, Restrictive, and No license). The Mann-Whitney \textit{U} test is a non-parametric test that examines whether two independent samples come from the same distribution, which does not require the data to be normally distributed or to meet the assumption of homogeneity of variance \cite{mcknight_mann-whitney_2010}. Given the large sample sizes, the \textit{U} values are expected to be large, and the salient test statistic is the p-value which indicates the statistical significance of observed differences. Due to capacity constraints in labelling licenses, we limited this analysis to repositories with  licenses used in at least 100 repositories ($n=$339,502, 98\% of all repositories). Subsequently, we analysed a snapshot of the social network structure of collaboration on the HF Hub (\textbf{RQ2}), using techniques defined in Table~\ref{tab:HF-network-properties} in Appendix~\ref{HF-appendix0}. This analysis provides insights into collaboration patterns in model repositories at this point in time. Furthermore, we analysed collaboration patterns in the three AI sub-fields (NLP, CV, and MM) to enable comparisons. Lastly, we examined model adoption on the HF Hub (\textbf{RQ3}) by calculating the ranked degree of models in the bipartite model usage networks and ranked degree of models in the model co-usage networks to identify the most used models in spaces and their respective developers. These two complementary approaches quantified model popularity (i.e. which models are most frequently used in spaces) and model co-popularity (i.e. which models are most commonly used in conjunction with other models). We replicated this analysis for spaces with NLP, CV, and MM tags for comparative analysis of the three AI sub-fields. 

\section{Results}

We first report results for activity in the 348,181 model, 65,761 dataset, and 156,642 space repositories in Section~\ref{HF-results-1}, relying on the metrics described in Section~\ref{HF-datacollection}. We then report results on the structure and dynamics of collaboration in Section~\ref{HF-results-2}, based on the analysis of collaboration interactions between around 100,000 developers in model repositories. Finally, we present the results of our analysis of model adoption in spaces in Section~\ref{HF-results-3}, where we examine the distribution of model usage in spaces on the HF Hub and identify the developers of the most used models.

\subsection{Development Activity on the HF Hub} \label{HF-results-1}
In this section, we present the findings of development activity in the repositories of 348,181 models, 65,761 datasets, and 156,642 spaces on the HF Hub. We present three key findings: right-skewed distributions across different types of activity (Section~\ref{HF-RQ1-rightskeweddist}), strong correlations between development activities (Section~\ref{HF-RQ1-correlations}), and a significant lack of licenses in model and dataset repositories (Section~\ref{HF-RQ1-licenses}).

\subsubsection{Right-Skewed Distributions in Development Activity}\label{HF-RQ1-rightskeweddist}
\begin{figure*}[t]
  \centering
\includegraphics[width=1\textwidth,height=0.68\textheight]{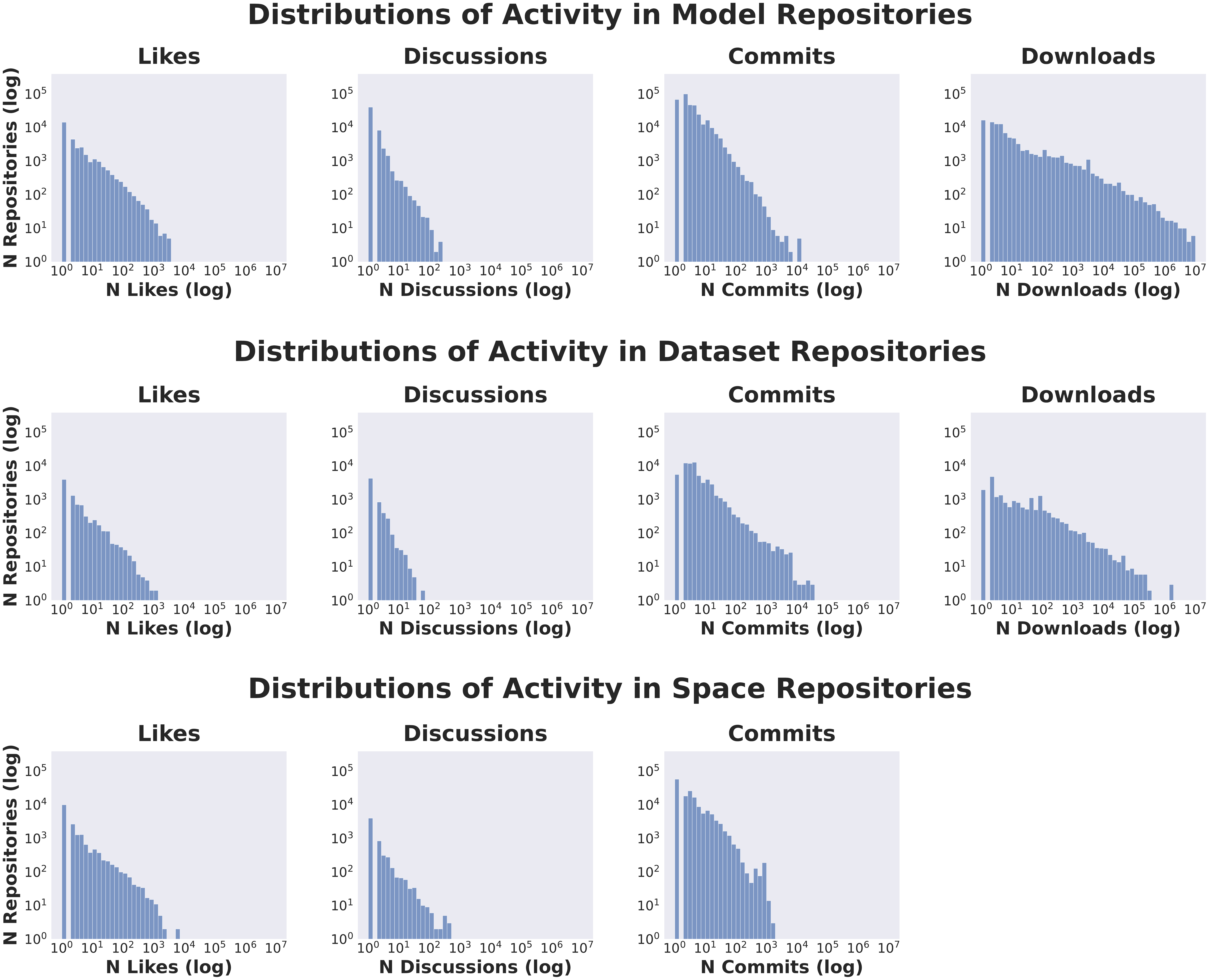}
  \caption{Distributions of Development Activity in HF Hub Repositories}
  \label{fig:HF-model-distribution}
\end{figure*}

Activity per repository is extremely imbalanced, with right-skewed distributions of \texttt{n\_likes}, \texttt{n\_discussions}, \texttt{n\_commits}, and \texttt{n\_downloads} across model, dataset, and space repositories (see Figure \ref{fig:HF-model-distribution}). For example, while the maximum number of likes amongst models is over 9,000, the average model only receives 1.14 likes (see Tables~\ref{tab:HF-models-summarytable}-\ref{tab:HF-spaces-summarytable}). The majority of repositories get minimal engagement. For example, 91\% of models and 88\% of datasets have 0 likes; 84\% of models, 91\% of datasets, and 96\% of spaces have 0 discussions; and 71\% of models and 70\% of datasets have 0 downloads. Meanwhile, most activity is concentrated in a small number of repositories. For example, $<1$\% of models account for 80\% of likes, 10\% for 80\% discussions, 30\% for 80\% commits, and $<1$\% for 80\% downloads. Upon increasing the threshold, 8\% of models account for 99\% likes, 15\% for 99\% discussions, and 1\% for 99\% downloads. 

Most repositories have a community size of 1; for example, 87\% of model repositories have 1 contributor and the 75\textsuperscript{th} quartile value of  \texttt{n\_committers} is 1 across repository types (see Table~\ref{tab:HF-models-summarytable}). The respective maximum values of \texttt{n\_committers} are 18, 100, and 282 across repository types, and the respective maximum values of \texttt{n\_community} are 246, 110, and 4,685. The differences between \texttt{n\_committers} and \texttt{n\_community} are due to large \texttt{n\_disc\_starters} values, indicating a division of roles in repositories, where many developers participate in discussions but few are involved in model maintenance. The model repositories with the most \texttt{n\_committers} are \texttt{bigscience/bloom} ($n$=$18$), \texttt{bigcode/santacoder} ($n$=$16$), and \texttt{deepset/roberta-base-squad2} ($n$=$15$). 

\subsubsection{Correlation between Community Size and Engagement} \label{HF-RQ1-correlations}
\begin{figure}[t]
  \centering
  \includegraphics[width=\textwidth,height=0.29\textheight]{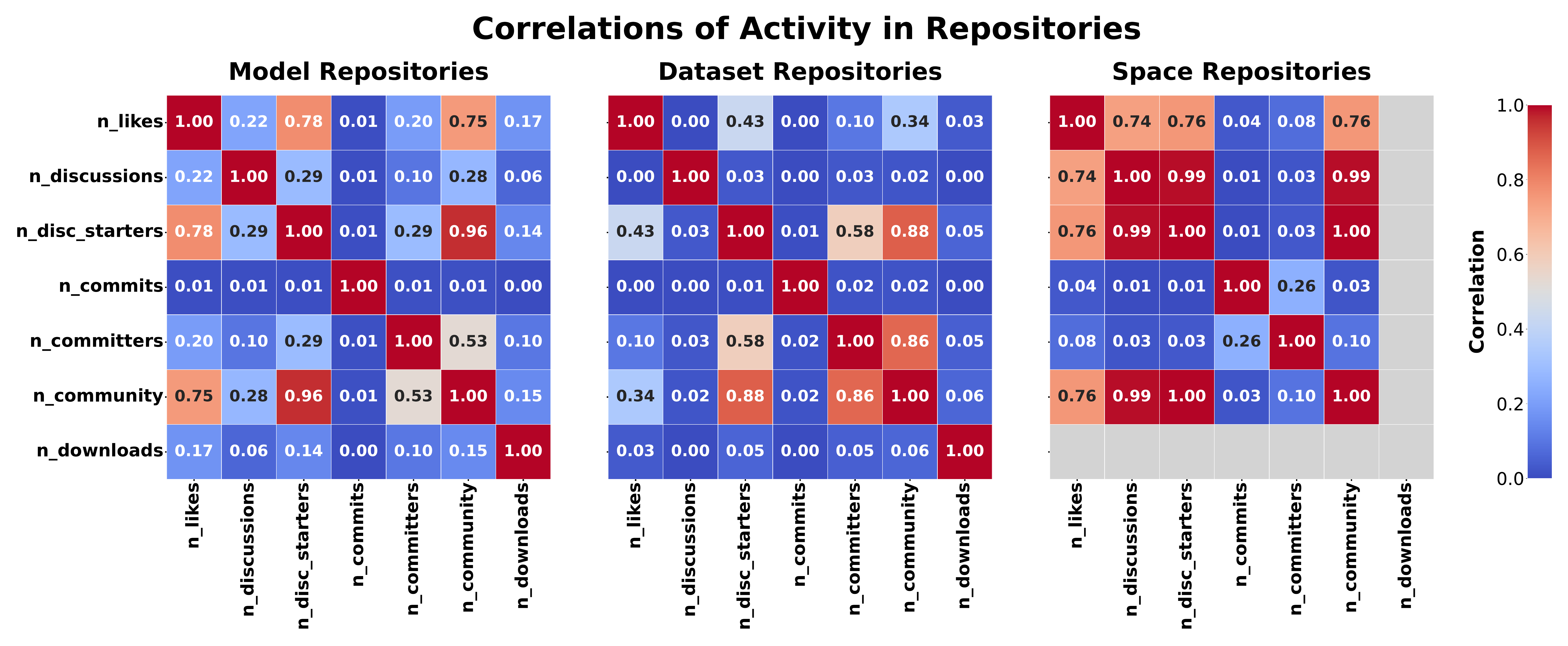}
  \caption{Correlations of Activity in Model, Dataset, and Space Repositories}
  \label{fig:HF-correlation-heatmap}
\end{figure}

We correlate frequency counts over the different types of activity described in Section~\ref{HF-datacollection} (see Figure~\ref{fig:HF-correlation-heatmap}). In model repositories, we find a strong positive correlations between \texttt{n\_community} and \texttt{n\_likes} ($\rho = 0.75$, $p < 0.001$). In space repositories, we find strong correlations between various activities, especially \texttt{n\_likes}) and \texttt{n\_discussions} ($\rho = 0.74$, $p < 0.001$), \texttt{n\_disc\_starters} ($\rho = 0.76$, $p < 0.001$), and \texttt{n\_community} ($\rho = 0.76$, $p < 0.001$). However, in general, we observe weak correlations between most activities in model and dataset repositories. Furthermore, we do not find a strong correlation between commit activity (\texttt{n\_commits}) and other types of activity, indicating that commit activity is not strongly linked to community engagement.

\subsubsection{Impact of Licenses on Collaboration} \label{HF-RQ1-licenses}

A significant proportion of model and dataset repositories lack licenses, which can create uncertainty and potential legal issues for users and developers. Specifying a license is not the norm: the majority of model repositories (65\%) and datasets (72\%) do not have a license. Amongst the licensed models, the most commonly used licenses are Apache v2.0 (37\%), MIT (17\%), OpenRAIL (14\%), and CreativeML OpenRAIL-M (10\%). The most used licenses for datasets are MIT (28\%), Apache v2.0 (15\%), OpenRAIL (9\%), and licenses from the family of Creative Commons v4.0 (7\%). 

\vspace{1em}

\begin{figure}[h]
  \centering
  \includegraphics[width=\textwidth,height=0.29\textheight]{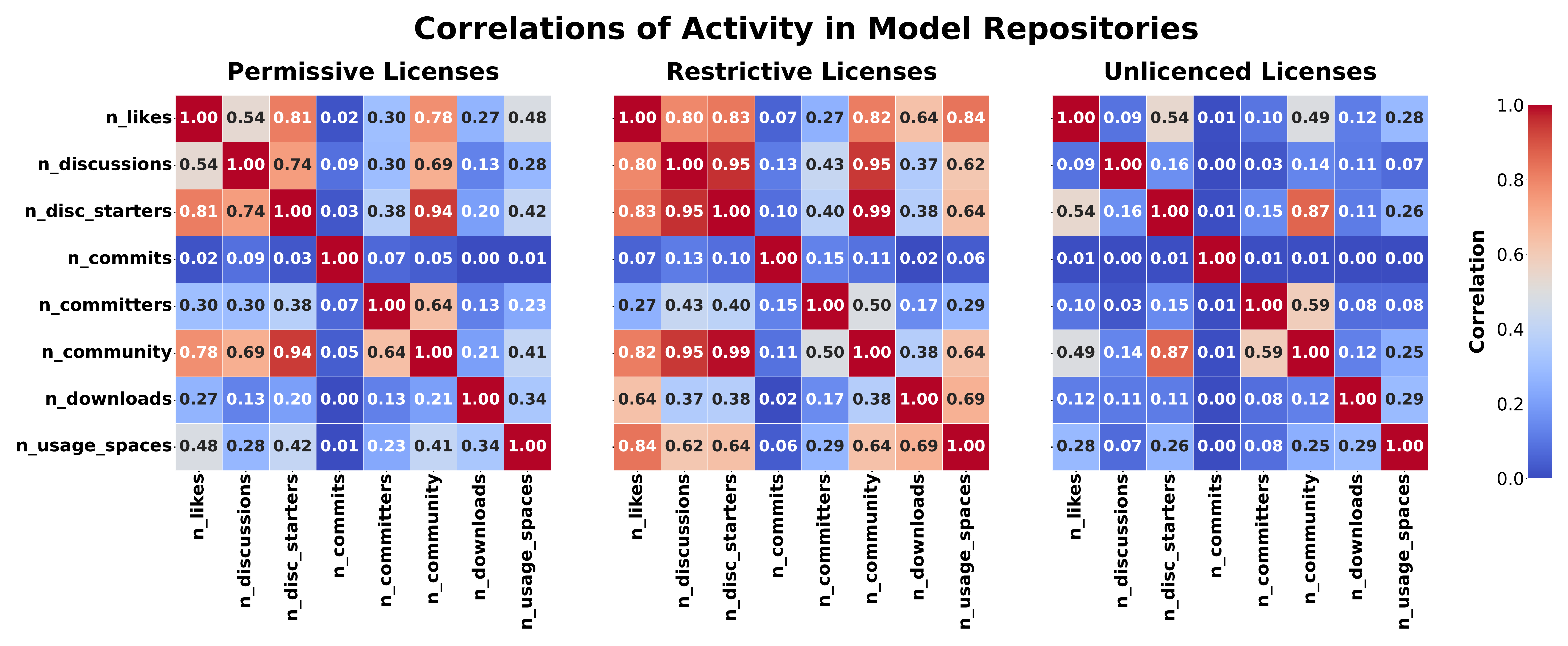}
  \caption{Correlations of Activity in Model Repositories with Different Licenses}
  \label{fig:HF-correlation-heatmap-licenses}
\end{figure}

The choice of license matters: there is a moderate to strong correlation between the use of a license and level of activity in model repositories (see Figure~\ref{fig:HF-correlation-heatmap-licenses}). Furthermore, the Mann-Whitney \textit{U} tests provide strong evidence of statistically significant differences between collaboration dynamics in model repositories with different types of licenses (all tests have $p < 0.001$). Specifically, model repositories with permissive licenses consistently have the highest levels of activity compared to model repositories with no license and those with restrictive licenses (see Table~\ref{tab:HF-collaboration-licenses}). However, repositories with restrictive licenses also exhibit significantly higher activity than those with no license. This pattern holds across all activity metrics measured, suggesting that while permissive licenses foster the highest engagement, restrictive licenses also promote more collaboration compared to model repositories that do not have a license.

\subsection{Social Network Structure and Dynamics of Collaboration} \label{HF-results-2}
In this section, we present findings from our analysis of a snapshot of the social network structure of collaboration in model repositories on the HF Hub. We begin with the structure and dynamics of collaboration in all model repositories (see Section~\ref{HF-results-collabnet}), and then we compare collaboration patterns in Natural Language Processing (NLP), Computer Vision (CV), and Multimodal (MM) model repositories (see Section~\ref{HF-results-collabnet-subfields}). 

\subsubsection{Collaboration in Model Repositories on the HF Hub} \label{HF-results-collabnet}

\begin{figure*}[t]
\centering
\includegraphics[width=1\textwidth,height=0.3\textheight]{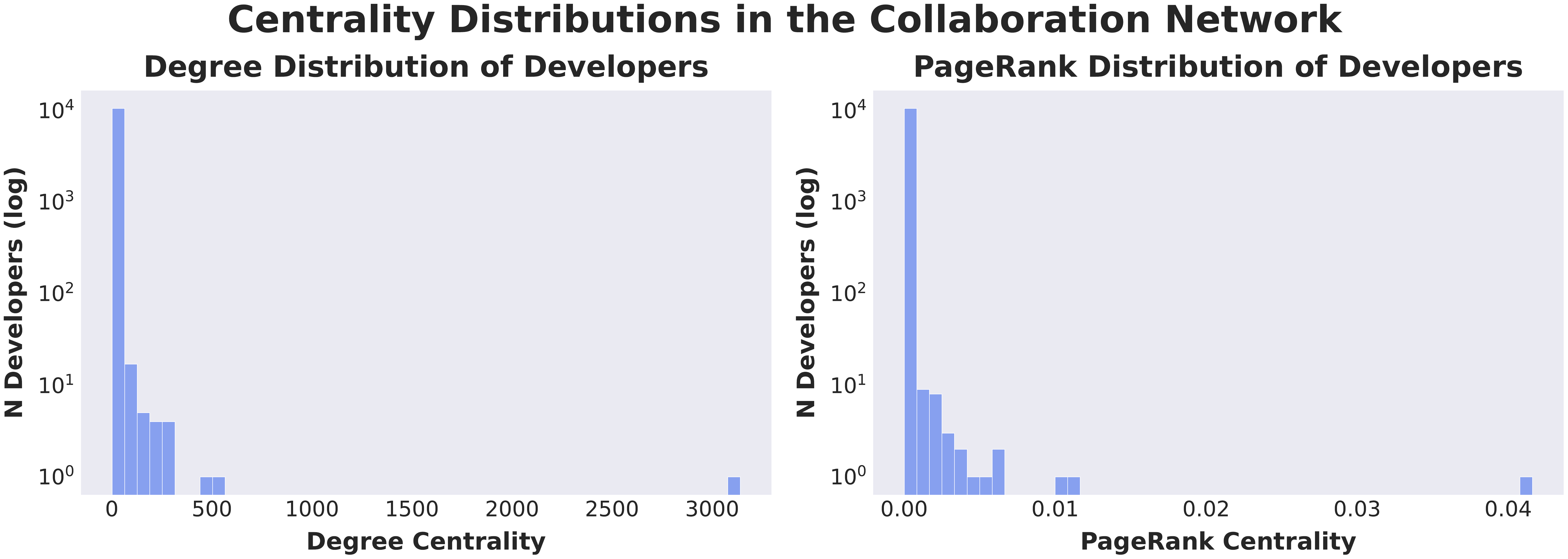}
  \caption{PageRank and Degree Distributions of Developers on the HF Hub}
  \label{fig:HF-collaboration-degreedistributions}
\end{figure*}

The HF Hub collaboration network exhibits a right-skewed degree and PageRank centrality distributions, which indicates that influence in the HF developer community is concentrated amongst a small subset of developers. The majority of developers (89\%) have not collaborated with others. Excluding these isolate developers, the remaining 10,524 developers have an average degree of 4.10 (SD: 32.63) and node degrees range from 1 to 3,140. The right-skewed distributions of degree and PageRank centrality (see Figure~\ref{fig:HF-collaboration-degreedistributions}) suggest that a small group of influential developers plays a central role in driving collaboration on open models on the HF Hub. Specifically, the degree centrality distribution has a mean of 4 and a median of 2, with a maximum of 3,140 and a standard deviation of 33, while the PageRank centrality distribution has a mean and median of 0.0001, a maximum of 0.04, and a standard deviation of 0.0005.

The HF Hub developer community exhibits a core-periphery structure, with a tightly interconnected core of prolific developers. The \textit{k}-core decomposition analysis reveals that as the \textit{k}-core value increases, the number of distinct communities decreases, ultimately converging into a single densely interconnected core at \textit{k}=26 (see Table~\ref{tab:HF-collab-structure}). The high modularity (0.81) at \textit{k}=1 suggests that the whole network consists of loosely connected groups of developers. As the \textit{k}-core value increases, the modularity decreases to 0.00 at \textit{k}=26, indicating a transition from a compartmentalised community structure with distinct clusters or modules to an integrated core characterised by high cohesion and a lack of discernible sub-groups. Concurrently, the sub-network density increases, reaching unity at \textit{k}=26.

Collaboration is characterised by high reciprocity values, ranging from 0.81 to 1.00 across all \textit{k}-core levels (see Table \ref{tab:HF-collab-structure}), indicating the prevalence of mutual relationships amongst developers. The low assortativity values, ranging from -0.49 to 0.08, suggest that developers collaborate regardless of their centrality in the network, implying that other factors, such as shared interests, skills, or project roles, may be more significant in driving collaboration than their network centrality. Furthermore, the relatively low average rich club coefficients, ranging from 0.04 to 0.41, indicate that highly central developers do not primarily collaborate with each other and a lack of elitism amongst power developers.


\subsubsection{Collaboration in Model Repositories in AI Sub-Fields} \label{HF-results-collabnet-subfields} 

Collaborations on models in sub-fields of natural language processing (NLP), computer vision (CV), and multimodal (MM), despite the different sizes of the respective communities, are similarly characterised by core-periphery structures ith high modularity and low density (see Tables~\ref{tab:HF-collab-structure-CV},\ref{tab:HF-collab-structure-NLP}, and \ref{tab:HF-collab-structure-MM}). At \textit{k}=1, all networks are highly modular (CV: 0.80, NLP: 0.82, MM: 0.71) and have very low density (CV: 0.01, NLP: 0.00, MM: 0.00), implying that collaborations in the respective AI sub-fields are clustered into distinct communities of collaborators. As the \textit{k} threshold increases, the networks undergo a similar transformation process, with modularity decreasing to 0.00 and the number of communities reducing to a single cohesive community at the maximal \textit{k} values (CV: 10, NLP: 25, MM: 26). Concurrently, density increases, reaching 1.00 for CV and MM and 0.97 for NLP at their respective maximal \textit{k} values. 

Collaboration in sub-fields is also similarly characterised by reciprocity and connectivity in the core. At \textit{k}=1, reciprocity values range from 0.84 to 0.93 and increase to 1.00 at the maximal \textit{k} for CV and MM, while NLP maintains a high reciprocity of 0.98 at its maximal \textit{k}. The average degree increases with \textit{k} for all networks, reaching the corresponding maximal \textit{k} value at the highest threshold. This suggests that as we move towards the core of the collaboration networks, developers become more interconnected and collaborate with a larger number of peers. However, the low average clustering coefficients and low average rich club coefficients across all networks indicate that the more prolific developers in the respective sub-fields tend to collaborate with a diverse set of individuals rather than forming tightly-knit groups. 

\vspace{1em}

\subsection{Model Adoption in Spaces on HF Hub} \label{HF-results-3}
In this section, we present the results of the analysis of model usage in spaces on the HF Hub, shedding light on model adoption and key developers in this ecosystem. Specifically, we present two key findings: model adoption in spaces is characterised by a right-skewed distribution (Section~\ref{HF-RQ3-rightskeweddistribution}), and a small cohort of developers (in particular, Big Tech companies) build the most used models across all spaces as well as in the three AI sub-fields (Section~\ref{HF-RQ3-developers}).

\subsubsection{Right-Skewed Distribution of Model Adoption} \label{HF-RQ3-rightskeweddistribution}

\begin{figure*}[t!]
  \centering
\includegraphics[width=1\textwidth,height=0.3\textheight]{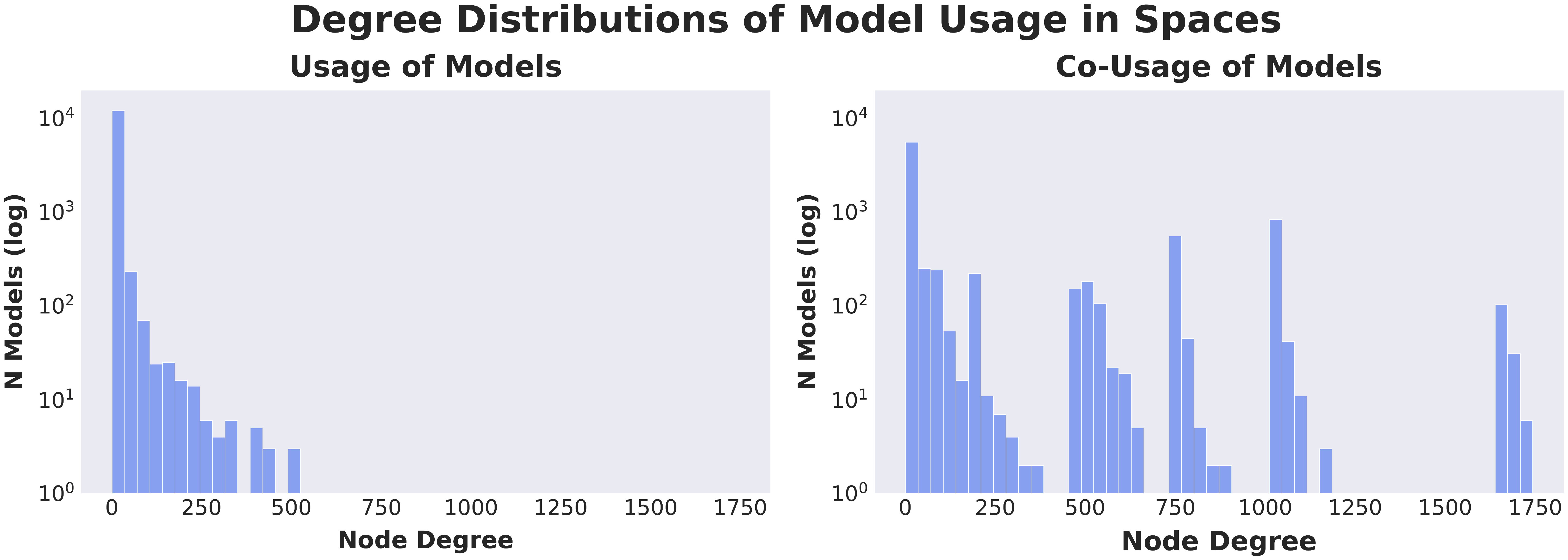}
\caption{Degree Distribution of Model Adoption in Spaces on the HF Hub}
  \label{fig:HF-modeldependency-distribution}
\end{figure*}

The bipartite model usage network displays a disparity in model adoption in spaces. The degree distribution of the bipartite network is right-skewed, as shown in Figure~\ref{fig:HF-modeldependency-distribution}. Only three models are used in 1,000 or more spaces, including \texttt{runwayml/stable-diffusion-v1-5}  ($n$=$1747$), \texttt{skytnt/anime-seg}  ($n$=$1162$), and \texttt{gpt2} ($n$=$1002$). The mean degree (6.68) is significantly higher than the median (1.00), and the large standard deviation (34.75) confirms the high variability in model usage. The majority of models have a low degree of usage, with at least 50\% being used in only one space, while a small number of highly popular models dominate the usage, with the maximum degree reaching 1,747. This suggests that a few key models are widely adopted in AI applications, while many other models have limited use cases. The model co-usage network provides an additional perspective on the uneven interdependencies of models in spaces, complementing insights gained from examining model downloads or individual model usage in spaces. Specifically, the degree distribution of this network exhibits a multi-modal pattern, with five distinct clusters, each exhibiting a right-skewed shape (see Figure~\ref{fig:HF-modeldependency-distribution}). A small cluster at the far-right tail of the distribution represents a few highly interconnected models with significantly higher co-usage degrees compared to the other clusters.

\subsubsection{Dominance of a Few Models by a Few Developers} \label{HF-RQ3-developers}

When we rank the models by their usage in spaces, we observe that major organisations, rather than individual developers or grassroots initiatives, have developed the most used models. Amongst the 100 most used models in spaces, the following organisations have developed the most models: Meta ($n$=$8$), Google ($n$=$7$), StabilityAI ($n$=$5$), OpenAI ($n$=$4$), Microsoft ($n$=$4$), and Fudan University ($n$=$4$). These five organisations account for 33\% of the 100 most used models in spaces. We note that the individual user nitrosocke ($n$=$5$), an employee at StabilityAI, ranked highly amongst these organisations. With regards to the model co-usage network, the key developers of the 100 most co-used models in all spaces are: EleutherAI ($n$=$15$), Meta ($n$=$12$), h20ai ($n$=$11$), BigScience ($n$=$9$), and lmsys ($n$=$9$). These five organisations account for 56\% of the 100 most co-used models in spaces. 

The model usage networks in the sub-fields similarly exhibit right-skewed degree distributions, highlighting the dominance of a minority of models in each sub-field. The most used models in spaces with NLP tags ($n$=$3,995$) are \texttt{gpt2} ($n$=$1,001$), \texttt{bert\-base\-uncased} ($n$=$621$), and \texttt{gpt2\-medium} ($n$=$445$). The organisations that developed the most models amongst the 100 most used models are Google ($n$=$9$), Meta ($n$=$5$), and Fudan University ($n$=$5$). For comparison, in the NLP model co-usage network, EleutherAI ranks first ($n$=$16$), followed by h20ai ($n$=$12$) and Meta ($n$=$11$).  The most used models in spaces with CV tags ($n$=$416$) are \texttt{saltacc/anime-ai-detect} ($n$=$500$), \texttt{openai/clip-vit-large-patch14} ($n$=$454$), and \texttt{openai/clip-vit-base-patch32} ($n$=$277$). The most prolific developer of models in spaces with CV tags is the user lllyasviel ($n$=$20$), followed by Meta ($n$=$8$) and the user DucHaiten ($n$=$7$). For comparison, in the CV model co-usage network, LAION AI ranks as the developer of the most models amongst the top 100 ($n$=$17$). Finally, the most used models in spaces with MM tags ($n$=$2,394$) are \texttt{runwayml/stable-diffusion-v1-5} ($n$=$1748$), \texttt{CompVis/stable-diffusion-v1-4} ($n$=$925$), and \texttt{stabilityai/stable-diffusion-2-1} ($n$=$854$). Amongst the developers of the 100 most used models, Stability AI ranks first, with 15 of the 100 most used models and 22 of the top-ranked co-used models in such spaces. These findings highlight the key models and players in the NLP, CV, and MM communities.

\subsubsection{Correlations between Model Likes and Model Usage}
We observe a strong positive correlation between \texttt{n\_likes} of models and \texttt{n\_usage\_spaces} ($\rho = 0.66$, $p < 0.001$), and a weak positive correlation between \texttt{n\_downloads} and \texttt{n\_usage\_spaces} ($\rho = 0.29, p < .001$). These findings suggest that the number of likes is more strongly associated with the usage of models in spaces compared to the number of downloads, and that likes in model repositories are a good indicator of its adoption in applications on the HF platform. However, as mentioned in Section~\ref{HF-datacollection}, we note that download counts are limited and therefore may only provide a snapshot of correlations between downloads and likes or usage, which may not generalise in all time periods.

\section{Discussion}

In this section, we discuss the key implications of our findings for research and practice. We highlight the study's contributions to the literature in Section \ref{HF-implications-research-contributions}. We then reflect on the methodological considerations of using the HF Hub as a data source for research on open source AI in Section \ref{HF-implications-research-methods}. Building on these insights, we make five recommendations for future research to advance the research agenda on open source AI in Section \ref{HF-implications-research-suggestions}. Finally, we discuss the implications for practice and make recommendations to practitioners in Section \ref{HF-implications-practice}.

\subsection{Implications for Research} \label{HF-implications-research}

\subsubsection{Contributions to Academic Literature} \label{HF-implications-research-contributions}

\textbf{Uneven influence in the HF Hub developer community:} We extend prior findings of right-skewed distributions of commit activity in model repositories \cite{castano_analyzing_2024} with observations of right-skewed distributions of various development activities on the HF Hub, including interactions in model, dataset, and space repositories; code collaborations between developers; and model usage in spaces. Activity distributions follow power law patterns, with a small fraction of repositories accounting for most interactions (e.g., $<1$\%  for 80\% of likes, 10\% for 80\% discussions, 30\% for 80\% commits, <1\% for 80\% downloads). Similarly, the collaboration networks exhibit right-skewed centrality distributions, indicating that influence is concentrated amongst few developers, congruent with prior observations that OSS development patterns generally follow Pareto distributions \cite{goeminne_evidence_2011,mockus_two_2002,szymanski_applicability_2023,xu_12_2006,zhang_companies_2021}. Influence also flows across the HF Hub, with likes per model having strong correlations with their usage in spaces ($\rho = 0.66$, $p < 0.001$).

\textbf{Impact of license on collaboration:} The Mann-Whitney \textit{U} tests show that license choice significantly impacts the level of activity and engagement in repositories, with permissive licenses exhibiting the highest activity levels, followed by repositories with restrictive licenses, and finally ones with no license. Furthermore, the Pearson correlations indicate that the use of a license (permissive or restrictive) is associated with stronger correlations between various types activity compared to repositories without licenses. These findings highlight the important role of licencing decisions in influencing the collaborative and community dynamics in open model development and open source AI projects.

\textbf{Core-periphery structure of the HF developer community:} 
To the best of our knowledge, only one prior study has investigated model development practices in the HF developer community, showing that most models only have one contributor and that model maintenance chiefly involves ``perfective tasks'' to enhance model performance \cite{castano_analyzing_2024}. We extend this finding with three insights. First, we corroborate the findings that most developers (89\%) are islands, who have not collaborated with other developers in model repositories on the HF Hub. This is not unique to the HF Hub: the majority of OSS projects are developed by individuals \cite{krishnamurthy_cave_2005}. However, what may be specific about the small community sizes in model development is the nature of the model development life-cycle (``code once, train often''). Second, the social network structure of collaboration patterns amongst developers in model repositories is characterised by a core-periphery structure, with a dense core of highly active developers, akin to the ``layered onion’’ structure common in OSS \cite{crowston_effective_2005}. 
Third, collaborations have high reciprocity and low assortativity, signifying the prevalence of mutual relationships amongst developers, regardless of their social positions in the community. 

\textbf{Uneven model adoption in spaces:} By examining model adoption in spaces, we empirically tested the observation of uneven model adoption and the disproportionate influence of industry-leading companies in the open source AI ecosystem \cite{widder_open_2023}. 
We identified the popularity of a relatively small number of models used in spaces as well as the influential role of a few organisations, including Meta, Google, Stability AI, OpenAI, Microsoft, and EleutherAI, who have developed the most widely used models. Some critics of the open-source model of AI development fear that too many unknown actors will introduce distributed safety issues, while advocates of the development model tout democratisation of power as a core benefit. Our findings show that a few organisations possess majority influence in this ecosystem, which challenges both of these narratives. In many cases, the most influential actors in the open source AI ecosystem are one in the same as those in closed-source AI \cite{widder_open_2023}. 

\subsubsection{HF Hub: A New Source of Research Data} \label{HF-implications-research-methods}
This paper contributes to the research effort to use the HF Hub as a data source for empirical studies on open model development \cite{ait_suitability_2023, ait_hfcommunity_2023,castano_analyzing_2024}. We share two reflections on methodological considerations. First, informed by prior work that underlines the importance of merging usernames for unique developers, we anticipated that this might be an issue on the HF Hub \cite{bird_mining_2006,goeminne_comparison_2013,kouters_whos_2012,robles_developer_2005,zhang_companies_2021}. While our three-pronged approach strikes a balance between the impracticality of manually inspecting over 100,000 developers versus the risk of misclassification through a fully automated approach, it is still imperfect. Future research may consider more sophisticated approaches to this problem.  Second, the API is not optimised for research purposes, which makes data collection time-consuming (e.g., one must make a unique API call to retrieve commit histories of each model and handle rate limits) and limited (e.g., user metadata is not available). The lack of user metadata hinders the ability to study the characteristics and behaviours of individual developers, such as their expertise and affiliations, as well as automated approaches to username merging that incorporate user metadata. To overcome these limitations, researchers may explore alternative approaches and tools, such as the HFCOMMUNITY database developed by Ait et al. to facilitate empirical studies of activity on the platform \cite{ait_hfcommunity_2023}.

\subsubsection{Recommendations for Future Research} \label{HF-implications-research-suggestions}
We recommend five research directions that can advance the research agenda on open source AI.

\textbf{1. Implications of concentrations in the HF Hub developer community:} We confirm prior observations that the models of a handful of companies are dominant amongst the HF Hub developer community \cite{widder_open_2023}. We encourage future research to investigate what these concentrations mean in practice, such as the potential benefits that these companies accrue from their open model ecosystems, including increased visibility, crowdsourced contributions (e.g., via commits and discussions), and access to diverse fine-tuned versions shared by other developers on the HF Hub. Furthermore, there is a concern that dominant companies benefit from developers being locked-in to their ecosystems, potentially limiting competition and entrenching their dominance. Future research could investigate the factors contributing to such concentrations, such as the reputation of the companies developing the models, their access to resources and support, or the perceived performance and versatility of their models, as well as the implications of these concentrations for the broader AI community, including the impact on research, innovation, and the distribution of benefits and resources.

\textbf{2. Incentives and modes of participation:} Future research could investigate the incentives of individual developers and companies. A number of companies have released open models on the HF Hub, such as Meta's LlaMA models \cite{huggingface_meta_2024}, Mistral AI's Mixtral models \cite{huggingface_mistral_2024}, and OpenAI's Whisper models \cite{huggingface_openai_2024}. Often these releases are presented as acts of ``AI democratisation'' \cite{seger_democratising_2023}. Future research could critically examine the commercial incentives behind these releases. In addition, future research could examine commercial approaches to model governance and maintenance---for example, if and how companies welcome or engage with community contributions---and if and how companies collaborate with each other on open model development, as they do in OSS development \cite{nguyen-duc_software_2019,germonprez_open_2013,linaker_how_2016, teixeira_collaboration_2014,zhang_how_2020}. 

\textbf{3. Collaboration dynamics in active repository communities:} We know that model maintenance focuses on model performance improvements \cite{castano_analyzing_2024}; and in the minority of repositories that have active communities, most developers contribute to discussions rather than commits (see Tables~\ref{tab:HF-models-summarytable}-\ref{tab:HF-spaces-summarytable}). Going further, we encourage researchers to examine collaboration dynamics in repositories with active communities from multiple angles. Given the sizeable differences in \texttt{n\_committers} and \texttt{n\_disc\_starters}, future research could investigate the division of roles between discussion and code contributors, typical topics of discussion (e.g., model performance, new ideas, etc.), how discussions inform model maintenance if at all, and the journeys of developers from discussion contributors to committers, amongst others. In addition, future research could examine the governance approaches (e.g., contribution policies) that repository owners use to encourage collaboration. Future analyses could also take into account temporal dynamics, providing insights into evolving patterns, social structures, and trends of open model developer communities on the HF Hub.

\textbf{4. Impact of model size on collaboration} Future research should examine the impact of model size (i.e., parameters) on the nature of collaboration in repositories on the HF Hub. For instance, it could examine how resource constraints (e.g., computational power or data availability) influence collaboration for various stakeholders (e.g., individual developers or developers from industry labs) on models of different sizes. By shedding light on facilitators and barriers for collaboration on open models, such research could guide efforts to foster inclusive and diverse communities.

\textbf{5. Collaboration beyond the HF Hub:} While this analysis provides insights into the developer community that shares and fine-tunes models on the HF Hub, we have a limited understanding of the development of the various components involved in the development of models \cite{white_model_2024}, which largely takes place in proprietary settings or on other platforms like GitHub \cite{ding_towards_2023}. We encourage future research to examine how the HF Hub is used in the wider ecosystem of platforms and offline venues for the collaborative development of open models and datasets. This research direction would enable comparisons of the collaboration patterns amongst model developers and model fine-tuners.  In addition, researchers could undertake a multi-sited analysis, examining collaboration on the same project across platforms. 

\subsection{Implications for Practice} \label{HF-implications-practice}

\subsubsection{Recommendations for Open Source Practitioners}
Beyond our academic research suggestions, we encourage open source researchers and practitioners to develop standardised metrics for studying open model development. Groups like the Linux Foundation's Community Health Analytics in Open Source Software (CHAOSS) working group \cite{chaoss_community_2024}, which has created metrics to assess the health and sustainability of OSS developer communities, are well-positioned to lead this effort. The lack of empirical data on open model development hinders evidence-based decision-making in this rapidly evolving field, and by working together to establish appropriate metrics, open source practitioners can help to address the data gap in ``open source AI''.

\subsubsection{Recommendations for Platform Providers}
We make two recommendations to HF as a platform for open model development. First, HF could work with researchers to identify features and API improvements that would aid research efforts concerning open model development on its platform, building on efforts by members of the HF community, such as \texttt{Weyaxi/huggingface-leaderboard}. This collaboration could include collecting and publishing data on open model development patterns and collaboration, which would help fill the current ``data gap'' in this area. HF may take inspiration from GitHub's Innovation Graph \cite{github_github_2024} or its annual Octoverse reports \cite{daigle_octoverse_2023}, which provide access to data and insights on development activity on its platform.  Second, a concerning proportion of models (64.67\%) and datasets (72.13\%) lack licenses, which may be due to uncertainty about how or whether they should be licensed \cite{hardy_should_2023,osi_deep_2023}. For comparison, the number of unlicensed repositories on GitHub is lower at 46\% or 53\% if including ``other licenses'' \cite{open_weaver_beware_2020}. In the interest of promoting responsible development, HF should consider developing educational resources on licenses, such as guides or tutorial videos, or developing features, such as a license drop-down menu, which can inform developers of the options available as well as their merits and drawbacks. Such a feature could be considered amongst other recommendations to moderate models on the HF Hub, such as hiring AI safety researchers and proactively red-teaming unsafe models \cite{tsamados_cybersecurity_2023,gorwa_moderating_2024}.  

\subsubsection{Recommendations for Policymakers} As open models become increasingly widely available and used, policymakers need empirical data to inform discussions about the benefits, risks, and governance of these models. Our analysis provides one empirical lens on the extent of model proliferation and adoption, which can help ground policy decisions. For example, it is illuminating to observe that most models (70.99\%) have not been downloaded once or that 1\% of models account for 99\% of downloads. This is a reminder that the availability of a model does not mean it will be (widely) used. Furthermore, while download counts were limited to the past 30 days, the fact that only 86 models had over one million downloads indicates that the number of widely used models is not excessively large and governable. What is more, the analysis revealed the impact of models developed by a number of non-profit, grassroots initiatives like EleutherAI, BigScience, and BigCode. Following the charge of the French government to fund the digital commons to support open source AI development \cite{chatterjee_france_2023}, policymakers may use such data to identify non-commercial projects that could be supported. Overall, the data points reported in Section~\ref{HF-results-1} could help policymakers assess the real-world impact of open models and develop appropriate governance frameworks to maximise their benefits while mitigating potential risks of open source AI. 

\subsection{Threats to Validity} \label{HF-threats-to-valdity}
We evaluate the validity of our findings by following guidance for empirical software engineering research \cite{easterbrook_selecting_2008,runeson_guidelines_2008}.

\subsubsection{Construct validity}
Construct validity concerns the extent to which a measurement accurately assesses the theoretical construct it intends to measure. Our study aimed to measure typical patterns of development activity on the HF Hub, but we acknowledge several threats to construct validity. First, our analysis is limited to activity in public repositories and does not account for collaboration in private repositories. Second, download counts have a few limitations: they are limited to the past 30 days, download counts may be incorrectly reported (e.g., if the repository lacks a configuration file or if the model is used on-device versus in continuous integration), and dataset downloads are limited to the count of \texttt{load\_dataset()} calls \cite{huggingface_models_2024,huggingface_datasets_2024}. Third, our operationalisation of collaboration relies on commits to model repositories, assuming that the co-occurrence of commits indicates collaboration. However, this assumption may not always hold true, especially in large repositories where developers may work on independent tasks. Future research could operationalise collaboration on specific files and quantify the relative contribution of developers to  specific files \cite{orucevic-alagic_network_2014}. Furthermore, this analysis is limited to snapshot of the HF Hub developer community in October 2023, which does not capture the dynamics of collaboration and activity over time, which should be considered in future research, as discussed in Section~\ref{HF-implications-research-suggestions}.

\subsubsection{Internal validity}
Internal validity concerns the extent to which a study can confidently attribute the observed results to the investigated variables, minimising the influence of confounding factors or alternative explanations. As explained in section \ref{HF-dataprocessing}, there may be a slight inaccuracy in the enumeration of community size per repository and the number of developers included in the collaboration networks due to discrepancies in username data, such as multiple accounts or usernames per developer. This is a common problem in OSS research, and there is no perfect solution to username merging \cite[]{bird_mining_2006,kouters_whos_2012,robles_developer_2005}. API limitations prevent the use of  methods that incorporate user metadata for username merging \cite{amreen_alfaa_2020,zhang_companies_2021}. For example, we rejected 34 username pairs due to insufficient evidence to confirm the match with confidence. 

\subsubsection{External validity}
External validity concerns the generalisability of the findings. While the HF Hub has gained significant popularity, it is important to acknowledge that there may be other platforms where open model development takes place and that our findings may not generalise to those platforms. Future research could explore collaboration practices  across different platforms to provide a more comprehensive view of the open source AI ecosystem. That being said, we observe that development activity on the HF Hub is characterised by the Pareto principle, conforming with  OSS development patterns on platforms like GitHub \cite{goeminne_evidence_2011,mockus_two_2002,szymanski_applicability_2023,xu_12_2006,zhang_companies_2021}. Another threat to the external validity of the findings concerns the analysis of model usage. While there were as many as 156,642 spaces at the time of data collection, they do not represent the use of open models beyond the HF Hub platform, thus limiting the generalisability of our claims, with the exception of finding a strong positive correlation between likes of model repositories and their usage in spaces ($\rho = 0.66$, $p < 0.001$). Future research could address this limitation by exploring other sources of data on model adoption, such as academic publications, industry reports, or user surveys, to triangulate the findings.

\subsubsection{Reliability}
Reliability refers to the consistency and reproducibility of the study's results. To enhance the reliability of our study, we have uploaded the Python scripts used for data collection and processing to a public GitHub repository \cite{osborne_python_2024}. Due to privacy and ethical considerations, we do not share the raw data (see Data Availability statement).

\section{Conclusion}
The burgeoning open source AI ecosystem has become a focal point of discussion amongst AI researchers, developers, and policymakers. This study offers empirical insights on practices in this emerging ecosystem via a quantitative analysis of development activity on the HF Hub. Concretely, we make three empirical contributions to the nascent research agenda on open source AI. First, we find that various types of development activity, from likes and downloads to discussions and commits, across 348,181 model, 65,761 dataset, and 156,642 space repositories exhibit right-skewed distributions. In addition, activity and engagement is highly imbalanced between repositories; for example, over 70\% of models have 0 downloads and 1\% account for 99\% of downloads. Second, we analyse a snapshot of the social network structure of collaboration in model repositories, finding that the community has a core-periphery structure, with a core of highly prolific developers and a majority of isolate developers (89\%) who do not collaborate with others. However, collaboration is characterised by high reciprocity and low levels of assortativity regardless of developers' social positions in the HF developer community. Third, we examine model adoption through the lens of model usage in spaces, finding that a minority of models are widely used and developed by a handful of industry-leading companies, which signifies the concentrated influence of a handful of actors in the HF Hub ecosystem. These findings are a timely reminder that open source AI is not immune to the influence of dominant industry leaders \cite{widder_open_2023}. We conclude with a discussion of the implications of our findings and recommendations for (open source) AI researchers, practitioners, and policymakers, with the hope that the practices in open model development can be more deeply investigated in the future. 

\section*{Acknowledgements}

The authors would like to thank Loubna Ben Allal, Daniel van Strien, Peter Cihon, Mer Joyce, Stefano Maffulli, Matt White, Seb Elmes, David Gray Widder, Alek Tarkowski, Johan Linåker, Sean P. Goggins, and the reviewers at the Journal of Computational Social Science for their generous feedback on prior versions of this manuscript.

\section*{Funding}

Cailean Osborne was supported by the Economic and Social Research Council Grant for Digital Social Science [ES/P000649/1]; Hannah Rose Kirk was supported by the Economic and Social Research Council Grant for Digital Social Science [ES/P000649/1]. Jennifer Ding was supported by the Ecosystem Leadership Award under the Engineering and Physical Sciences Research Council Grant [EP/X03870X/1] \& the Alan Turing Institute.

\section*{Data Availability Statement}
The authors may be contacted with enquiries about the research data, which may be shared upon reasonable request and only in compliance with applicable data protection regulations.

\section*{Conflict of Interests}
On behalf of all authors, the corresponding author states that there is no conflict of interest.

\bibliographystyle{unsrt}  
\bibliography{references}

\newpage
\appendix
\section{Definitions of Network Properties in Social Network Analysis} \label{HF-appendix0}

\begin{table}[h]
\centering
\footnotesize
\caption{Definition of Network Properties}
\begin{tabular}{|p{3cm}|p{9cm}|}
    \hline
       \textbf{Property} & \textbf{Definition} \\ \hline
        \texttt{Nodes} &  Number of nodes in the network (e.g., developers or models). \\ \hline
        \texttt{Edges} & Number of edges in the network (e.g., links between developers). \\ \hline

        \texttt{Degree centrality} & Degree centrality is a measure of the importance of a node in a network based on the number of connections it has. It is calculated as the number of edges a node has with other nodes. \\ \hline
        
        \texttt{PageRank centrality} &  PageRank is a network centrality measure that assesses a node's importance based on the quantity and quality of incoming links, considering the recursive influence of nodes pointing to it \cite{page_pagerank_1999}. PageRank values range from 0 to 1, with higher values indicating greater global influence and importance in the network. \\ \hline
        
        \texttt{k-core} &  \textit{k}-core decomposition identifies the maximal subgraph in which every node is connected to at least \textit{k} other nodes, helping to reveal the network's core structure \cite{batagelj_om_2003}. \\ \hline 
        
        \texttt{Modularity (MOD)} &  Modularity measures the extent to which a network can be divided into distinct and densely interconnected communities. The modularity value ranges from -1 to 1, with positive values indicating a high degree of community structure and negative values implying the absence of community structure \cite{newman_modularity_2006}. We used the Clauset-Newman-Moore greedy modularity maximization algorithm to calculate the modularity of networks \cite{clauset_finding_2004}. \\ \hline
        
        \texttt{Communities (COM)} & Community detection identifies cohesive groups in a network by optimizing a measure of modular structure. The goal is to find a partition of the network that maximizes the density of connections in communities while minimizing the connections between them. We used the Clauset-Newman-Moore greedy modularity maximization algorithm \cite{clauset_finding_2004} to find the community partition with the highest modularity value and to determine the number of communities in the network (\texttt{COM}). \\ \hline
        
        \texttt{Density (DENS)} &  Density is a measure of how connected a network is, calculated as the ratio of the number of edges present in the network to the maximum possible number of edges in the network \cite{networkx_density_2023}. \\ \hline
        
        \texttt{Reciprocity (RECIP)} &  The reciprocity of a directed graph is the ratio of the number of edges pointing in both directions to the total number of edges in the graph.\cite{networkx_reciprocity_2023}. A  value of 1 means that all edges are bidirectional, while a  value of 0 means there are no mutual connections. \\ \hline
        
        \texttt{Average Rich Club Coefficient (ARCC)} & The average rich club coefficient measures the extent to which high-degree nodes tend to be more connected than expected by chance \cite{zhou_rich-club_2004}. Detecting the rich-club phenomenon reveals high-level semantic insights about the network \cite{mcauley_rich-club_2007}, a key property of power-law networks \cite{zhou_rich-club_2004}. 
        The coefficient ranges from 0 to 1, where 0 indicates no preferential connection and 1 indicates a fully connected high-degree subgraph. We implemented this calculation on directed networks \cite{smilkov_rich-club_2010}. For each network, we set the threshold \textit{k} as the degree value corresponding to the minimum degree amongst the top 10\% of highest-degree nodes.
        \\ \hline
        
        \texttt{Assortativity (ASS)} &  Assortativity measures the tendency of nodes to be connected to nodes with similar degrees, with values ranging from [-1,1] \cite{newman_assortative_2002}. \\ \hline
        
        \texttt{Average Degree (AD)} & The average number of edges per node.\\ \hline
        
        \texttt{Average Clustering Coefficient (ACC)} & The average clustering coefficient [0, 1] measures the extent to which nodes in a network tend to cluster together. It quantifies the level of triadic closure in the network, which is the tendency for a node’s neighbours to be connected to each other \cite{saramaki_generalizations_2007}. ~ \\ \hline
    \end{tabular}
    \label{tab:HF-network-properties}
\end{table}

\newpage
\section{Summary Statistics of Development Activity in Repositories} \label{HF-appendix1}

\begin{table}[h]
\begin{minipage}{\textwidth}
\centering
\footnotesize
\caption{Summary Statistics of Development Activity in Model Repositories}
\begin{tabular}{|l|c|c|c|c|c|c|c|}
    \hline
         & \texttt{n\_likes} &   \texttt{n\_discussions} &\texttt{n\_disc\_starters} &\texttt{n\_commits} & \texttt{n\_committers}  & \texttt{n\_community}  &\texttt{n\_downloads} \\ \hline
        \textbf{mean} & 1.14 &   0.28 &0.10 &7.28 &  1.06 & 1.13  &1,693.93 \\ \hline
        \textbf{std} & 30.56 &   5.42 &1.17 &167.14 & 0.41 & 1.26  &158,207.09 \\ \hline
        \textbf{min} & 0 &   0 &0 &0 & 0 & 0  &0 \\ \hline
        \textbf{25\%} & 0 &   0 &0 &2 & 1 & 1  &0 \\ \hline
        \textbf{50\%} & 0 &   0 &0 &3 & 1 & 1  &0 \\ \hline
        \textbf{75\%} & 0 &   0 &0 &5 & 1 & 1  &0 \\ \hline
        \textbf{max} & 9,314 &   3,006 &240 &75,653 & 18 & 246  &65,729,394 \\ \hline
    \end{tabular}
    \label{tab:HF-models-summarytable}
    \textit{N.B. \texttt{n\_community} equals 0 if no user, apart from the repository creator, has made a commit or started a discussion in the repository.}
\end{minipage}

\vspace{0.5cm}

\begin{minipage}{\textwidth}
    \centering
    \footnotesize
    \caption{Summary Statistics of Development Activity in Dataset Repositories}
    \begin{tabular}{|l|c|c|c|c|c|c|c|}
    \hline
         & \texttt{n\_likes} &   \texttt{n\_discussions} &\texttt{n\_disc\_starters} &\texttt{n\_commits} & \texttt{n\_committers}  & \texttt{n\_community}  &\texttt{n\_downloads} \\ \hline
        \textbf{mean} & 0.92&   1.41&0.12&30.04&  1.13& 1.20&476.70\\ \hline
        \textbf{std} & 18.14&   309.14&0.61&1298.11& 0.67& 0.87&43786.43\\ \hline
        \textbf{min} & 0 &   0 &0 &0 & 0 & 0  &0 \\ \hline
        \textbf{25\%} & 0 &   0 &0 &2 & 1 & 1  &0 \\ \hline
        \textbf{50\%} & 0 &   0 &0 &4& 1 & 1  &0 \\ \hline
        \textbf{75\%} & 0 &   0 &0 &7& 1 & 1  &2\\ \hline
        \textbf{max} & 3493&   79,256&64&314,813& 100& 110&9,651,261\\ \hline
    \end{tabular}
    \label{tab:HF-datasets-summarytable}
    \textit{N.B. \texttt{n\_community} equals 0 if no user, apart from the repository creator, has made a commit or started a discussion in the repository.}
\end{minipage}

\vspace{0.5cm}

\begin{minipage}{\textwidth}
    \centering
    \footnotesize
    \caption{Summary Statistics of Development Activity in Space Repositories}
    \begin{tabular}{|l|c|c|c|c|c|c|c|}
    \hline
         &  \texttt{n\_likes} &   \texttt{n\_discussions} &\texttt{n\_disc\_starters}&\texttt{n\_commits} & \texttt{n\_committers}& \texttt{n\_community}  &\texttt{n\_downloads} \\ \hline
        \textbf{mean} & 1.33&   0.34&0.15&8.22&  1.27& 1.40&\texttt{N/A}\\ \hline
        \textbf{std} & 37.35 &   54.58&15.36&40.36& 0.99& 15.39& \texttt{N/A} \\ \hline
        \textbf{min} & 0 &   0 &0 &0 & 0 & 0  &\texttt{N/A}\\ \hline
        \textbf{25\%} & 0 &   0 &0 &1& 1 & 1  &\texttt{N/A}\\ \hline
        \textbf{50\%} & 0 &   0 &0 &3 & 1 & 1  &\texttt{N/A}\\ \hline
        \textbf{75\%} & 0 &   0 &0 &5 & 1 & 2&\texttt{N/A}\\ \hline
        \textbf{max} & 9,124&   18,061&4,684&2,150& 282& 4,685&\texttt{N/A}\\ \hline
    \end{tabular}
    \label{tab:HF-spaces-summarytable}
    \textit{N.B. \texttt{n\_community} equals 0 if no user, apart from the repository creator, has made a commit or started a discussion in the repository.}
\end{minipage}
\end{table}

\newpage
\section{Mann-Whitney U Tests for Activity in Model Repositories}\label{HF-appendix2}

\begin{table}[h]
\centering
\footnotesize
\caption{Mann-Whitney \textit{U} Test Results}
\begin{tabular}{|l|c|c|c|}
\hline
\textbf{Activity} & \multicolumn{1}{c|}{\textbf{Permissive vs.}} & \multicolumn{1}{c|}{\textbf{Permissive vs.}} & \multicolumn{1}{c|}{\textbf{Restrictive vs.}} \\
& \multicolumn{1}{c|}{\textbf{Restrictive}} & \multicolumn{1}{c|}{\textbf{No license}} & \multicolumn{1}{c|}{\textbf{No license}} \\
\hline
\texttt{n\_likes} & 1,464,356,547 & 8,712,552,137 & 5,558,887,829 \\
& ($p < 0.001$) & ($p < 0.001$) & ($p < 0.001$) \\
\hline
\texttt{n\_disc\_starters} & 1,501,884,124 & 8,323,015,321 & 5,200,616,824 \\
& ($p < 0.001$) & ($p < 0.001$) & ($p < 0.001$) \\
\hline
\texttt{n\_discussions} & 1,727,651,315 & 9,180,111,459 & 4,983,923,782 \\
& ($p < 0.001$) & ($p < 0.001$) & ($p < 0.001$) \\
\hline
\texttt{n\_committers} & 1,573,347,885 & 8,744,559,271 & 5,238,059,429 \\
& ($p < 0.001$) & ($p < 0.001$) & ($p < 0.001$) \\
\hline
\texttt{n\_commits} & 2,013,736,368 & 11,333,305,055 & 5,451,995,655 \\
& ($p < 0.001$) & ($p < 0.001$) & ($p < 0.001$) \\
\hline
\texttt{n\_downloads} & 1,884,325,690 & 10,136,574,180 & 4,983,179,068 \\
& ($p < 0.001$) & ($p < 0.001$) & ($p < 0.001$) \\
\hline
\texttt{n\_community} & 1,563,484,573 & 8,856,297,978 & 5,333,406,861 \\
& ($p < 0.001$) & ($p < 0.001$) & ($p < 0.001$) \\
\hline
\texttt{n\_usage\_spaces} & 1,509,785,617 & 8,276,280,946 & 5,148,816,091 \\
& ($p < 0.001$) & ($p < 0.001$) & ($p < 0.001$) \\
\hline
\end{tabular}
\label{tab:HF-collaboration-licenses}
\end{table}

\newpage
\section{Social Network Structure of Collaboration on HF Hub} \label{HF-appendix3}

\begin{table}[h]
\centering
\footnotesize
\caption{Network Structure of Collaboration in All Model Repositories}
\begin{tabular}{|l|c|c|c|c|c|c|c|c|c|c|}
\hline
\textbf{\texttt{k-core}} & \texttt{Nodes} & \texttt{Edges} & \texttt{MOD} & \texttt{COM} & \texttt{DENS} & \texttt{RECIP} & \texttt{ARCC} & \texttt{ASS} & \texttt{AD} & \texttt{ACC} \\ \hline
1 & 10,524 & 21,598 & 0.81 & 2,894 & 0.00 & 0.83 & 0.04 & 0.08 & 4.10 & 0.00 \\ \hline
5 & 1,109 & 7,849 & 0.60 & 68 & 0.01 & 0.81 & 0.08 & 0.01 & 14.16 & 0.00 \\ \hline
10 & 330 & 3,769 & 0.52 & 18 & 0.03 & 0.85 & 0.11 & 0.07 & 22.84 & 0.00 \\ \hline
26 (MAX) & 14 & 182 & 0.00 & 1 & 1.00 & 1.00 & 0.15 & NaN & 26.00 & 0.04 \\ \hline
\end{tabular}
\label{tab:HF-collab-structure}
\vspace{0.5cm}

\centering
\footnotesize
\caption{Network Structure of Collaboration in Computer Vision Model Repositories}
\begin{tabular}{|l|c|c|c|c|c|c|c|c|c|c|}
    \hline
        \texttt{k-core} & \texttt{Nodes} & \texttt{Edges} & \texttt{MOD} & \texttt{COM} & \texttt{DENS} & \texttt{RECIP} & \texttt{ARCC} & \texttt{ASS} & \texttt{AD} & \texttt{ACC} \\ \hline
        1 & 371 & 697 & 0.80 & 112 & 0.01 & 0.90 & 0.26 & 0.26 & 3.76 & 0.01 \\ \hline
        5 & 41 & 203 & 0.52 & 4 & 0.12 & 0.93 & 0.67 & -0.00 & 9.90 & 0.04 \\ \hline
        10 (MAX) & 6 & 30 & 0.00 & 1 & 1.00 & 1.00 & 0.40 & NaN & 10.00 & 0.16 \\ \hline
    \end{tabular}
    \label{tab:HF-collab-structure-CV}
\vspace{0.5cm}

\centering
\footnotesize
\caption{Network Structure of Collaboration in Natural Language Processing Model Repositories}
    \begin{tabular}{|l|c|c|c|c|c|c|c|c|c|c|}
    \hline
        \texttt{k-core} & \texttt{Nodes} & \texttt{Edges} & \texttt{MOD} & \texttt{COM} & \texttt{DENS} & \texttt{RECIP} & \texttt{ARCC} & \texttt{ASS} & \texttt{AD} & \texttt{ACC} \\ \hline
        1 & 4,606 & 10,010 & 0.82 & 1,366 & 0.00 & 0.93 & 0.07 & 0.33 & 4.35 & 0.00 \\ \hline
        5 & 565 & 3,980 & 0.62 & 41 & 0.01 & 0.91 & 0.11 & 0.14 & 14.09 & 0.00 \\ \hline
        10 & 182 & 1,980 & 0.52 & 13 & 0.06 & 0.90 & 0.15 & 0.07 & 21.76 & 0.00 \\ \hline
        25 (MAX) & 15 & 204 & 0.00 & 1 & 0.97 & 0.98 & 0.25 & NaN & 27.20 & 0.01 \\ \hline
    \end{tabular}
    \label{tab:HF-collab-structure-NLP} 
\vspace{0.5cm}

\centering
\footnotesize
    \caption{Network Structure of Collaboration in Multimodal Model Repositories}
    \begin{tabular}{|l|c|c|c|c|c|c|c|c|c|c|}
    \hline
        \texttt{k-core} & \texttt{Nodes} & \texttt{Edges} & \texttt{MOD} & \texttt{COM} & \texttt{DENS} & \texttt{RECIP} & \texttt{ARCC} & \texttt{ASS} & \texttt{AD} & \texttt{ACC} \\ \hline
        1 & 1,546 & 3,661 & 0.71 & 321 & 0.00 & 0.84 & 0.10 & 0.19 & 4.74 & 0.00 \\ \hline
        5 & 218 & 1,480 & 0.57 & 18 & 0.03 & 0.89 & 0.16 & -0.00 & 13.58 & 0.00 \\ \hline
        10 & 82 & 805 & 0.50 & 6 & 0.12 & 0.91 & 0.37 & -0.33 & 19.63 & 0.02 \\ \hline
        26 (MAX) & 14 & 182 & 0.00 & 1 & 1.00 & 1.00 & 0.15 & NaN & 26.00 & 0.06 \\ \hline
    \end{tabular}
    \label{tab:HF-collab-structure-MM}
\end{table}

\end{document}